\documentclass[showpacs,preprintnumbers,superscriptaddress]{revtex4}
\usepackage{bbm}
\usepackage{mathrsfs}
\usepackage{CJK}
\usepackage{amsmath,amssymb,graphicx,bm}
\begin{document}
	\title{Accretion of matter by a Charged dilaton black hole}
	\author{Yinan Jia \footnote{Co-first author}}
	\affiliation{College of Physical Science and Technology, Hebei University, Baoding 071002, China}
\author{Tong-Yu He \footnote{Co-first author}}
\affiliation{College of Physics Science and Technology, Hebei University, Baoding 071002, China}

\author{Wen-Qian Wang}
\affiliation{College of Physics Science and Technology, Hebei University, Baoding 071002, China}

\author{Zhan-Wen Han}
\affiliation{College of Physics Science and Technology, Hebei University, Baoding 071002, China}
\affiliation{Yunnan Observatories, Chinese Academy of Sciences, Kunming 650216, China}

\author{Rong-Jia Yang \footnote{Corresponding author}}
\email{yangrongjia@tsinghua.org.cn}
\affiliation{College of Physics Science and Technology, Hebei University, Baoding 071002, China}
\affiliation{Hebei Key Lab of Optic-Electronic Information and Materials, Hebei University, Baoding 071002, China}
\affiliation{National-Local Joint Engineering Laboratory of New Energy Photoelectric Devices, Hebei University, Baoding 071002, China}
\affiliation{Key Laboratory of High-pricision Computation and Application of Quantum Field Theory of Hebei Province, Hebei University, Baoding 071002, China}

	\begin{abstract}
 Considering accretion onto a charged dilaton black hole, the fundamental equations governing accretion, general analytic expressions for critical points, critical velocity, critical speed of sound, and ultimately the mass accretion rate are obtained. A new constraint on the dilation parameter coming from string theory is found and the case for polytropic gas is delved into a detailed discussion. It is found that the dialtion and the adiabatic index of accreted material have deep
effects on the accretion process.
	\end{abstract}
	
	\maketitle
	\section{Introduction}
Accretion of matter onto a massive object is one of the most common processes in astrophysics. The groundwork for understanding accretion dates back to the early contributions of Hoyle, Lyttleton, and Bondi \cite{hoyle1939effect, 1940Obs6339L, Bondi:1944jm}. Bondi, in particular, pioneered Newtonian solutions that describe spherically symmetric accretion of a perfect fluid within the Keplerian gravitational potential \cite{Bondi:1952ni}. Subsequent to Bondi's work, Michel extended the model to General Relativity by considering spherical accretion of a perfect fluid onto a Schwarzschild black hole \cite{michel1972accretion}. The exploration of accretion processes has since been a subject of extensive investigation, with a plethora of scholarly works contributing to the field. Notable examples include studies in \cite{begelman1978accretion, Malec:1999dd, Babichev:2004yx, Babichev:2010kj, Rodrigues:2016uor, Contreras:2018gct, Abbas:2018ygc, Zheng:2019mem, Yang:2020bpj, UmarFarooq:2020aum, Nozari:2020swx, Panotopoulos:2021ezt, Iftikhar:2020ykp, Gao:2008jv, John:2013bqa, Jiao:2016iwp, Ganguly:2014cqa, Mach:2013fsa, Kremer:2020yfg, Tejeda:2019lie, Feng:2022bst, Rahaman:2022lpk}. The primary emphasis of these studies revolves around critical points, flow parameters, and accretion rates. In the study conducted by \cite{Yang:2015sfa}, the process of accretion onto a renormalization-group-improved Schwarzschild black hole was employed to examine the viability of the asymptotically safe scenario. This investigation delves into the intricate interplay between accretion dynamics and the improved gravitational features. Additionally, in the works of \cite{Jamil:2008bc} and \cite{Yang:2019qru}, the accretion conditions are explored, shedding light on constraints imposed on the mass-to-charge ratio. These analyses contribute valuable insights into the delicate balance between gravitational and electromagnetic forces governing the accretion processes in these diverse black hole scenarios.

In general, numerical analyses are essential for handling nonspherical accretion in both relativistic and Newtonian flows \cite{Papadopoulos:1998up, Font:1998sc, Zanotti:2011mb, Lora-Clavijo:2015hqa}. Therefore, it is crucial to seek analytical solutions for accretion processes. Exact, fully relativistic solutions were obtained in \cite{Petrich:1988zz}, describing the accretion of matter with an adiabatic equation of state onto a moving Schwarzschild or Kerr black hole. Building upon this work, analytic solutions were found for accretion onto a moving Kerr-Newman black hole \cite{Babichev:2008dy}, a moving Reissner-Nordstr{\"o}m black hole \cite{Jiao:2016uiv}, and a moving charged dilaton black hole \cite{Yang:2021opo}. In \cite{Liu:2009ts, Zhao:2018ani}, exact solutions were derived for shells accreted onto a Schwarzschild black hole. Recently, exact solutions were presented for the accretion of collisionless Vlasov gas onto a moving Schwarzschild black hole \cite{Mach:2021zqe, Mach:2020wtm} or a Schwarzschild-like black hole \cite{Cai:2022fdu} or a Kerr black hole \cite{Cieslik:2022wok}, based on the Hamiltonian formalism developed in \cite{Rioseco:2016jwc}. This method enables the analysis of more complex flows on a fixed background. However, due to computational complexity, obtaining the exact solution for mass accretion rate in the case of more complicated black holes remains challenging.

Here, our objective is to analyze in detail the accretion process for more general matter onto a charged dilaton black hole. Accretion onto this type of black hole has been studied in \cite{Yang:2021opo}, where the four-velocity of accreted flow, the particle number density, and the accretion rate were determined. The method adopted in \cite{Yang:2021opo} is only suitable for gaseous medium with a adiabatic equation of state ($p=\rho$), if we consider more general accreted matter, such as polytrope gas, we should employ the scheme presented in \cite{John:2013bqa, Jiao:2016iwp}. Secondly, if we want to investigate the temperature of accreted matter near horizon, the method used in \cite{Yang:2021opo} is also invalid. So it is necessary to reconsider analytically the critical points, the critical fluid velocity, the critical sound speed, the mass accretion rate, and subsequently the temperature near the horizon, and investigate the influences of model parameters on the accretion process. It is also interesting to compare the results here with those obtained in \cite{Yang:2021opo}.

The paper is structured as follows: In the next Section, we provide the fundamental equations governing the accretion of matter onto a charged dilaton black hole. We explores the critical points and the necessary conditions that these points must satisfy in Sect. III. We delves into a detailed discussion of the application involving polytropic gas in Sect. IV. Finally, we offer a concise summary and discussion of the findings in Sect. V.

	\section{BASIC EQUATIONS FOR ACCRETION}
The charged dilaton black hole is a solution within low-energy string theory which represents a static, spherically symmetric charged black hole \cite{Gibbons:1987ps, Garfinkle:1990qj}. The spacetime geometry is expressed by the following line element
   \begin{eqnarray}
		\label{1}
		ds^2=-\left(1-\frac{2M}{r}\right)dt^2+\left(1-\frac{2M}{r}\right)^{-1}dr^2+r(r-a)d\Omega^2,
	\end{eqnarray}
where $a=Q^2\frac{\exp{2\phi_{0}}}{M}$ with $\phi_0$ representing the asymptotic constant value of the dilaton. The parameter $M$ denotes the mass of the black hole as measured by an observer at infinity, and $Q$ signifies the electric charge of the black hole. Throughout this paper, we adopt a unit system in which the Newton's gravitational constant $G$ and the speed of light $c$ are set to 1. The event horizon of the black hole is located at $r_{H}=2M$. When $r=a$, the area of the sphere tends to zero, indicating a singularity on the surface. Since $g_{rr}>0$ and $r\geq 2M$, meaning that we have a limit on the parameter $a$: $a\leq 2M$. Taking $a= 2M$, the geometry \eqref{1} describes a 
naked singularity \cite{Garfinkle:1990qj}. 
	
	We can introduce a model of steady-state radial inflow of matter onto the charged dilaton black hole, neglecting back-reaction effects. The matter is modeled as an ideal fluid characterized by the following energy-momentum tensor
	\begin{eqnarray}
		\label{2}
		T_{\alpha\beta}=(\rho+p)u_{\alpha}u_{\beta}+pg_{\alpha\beta},
	\end{eqnarray}
where $\rho$ is the proper energy density and $p$ is the proper pressure for fluid. The four-velocity of fluid $u^{\alpha}=\frac{\mathrm{d}x^{\alpha}}{\mathrm{d}s}$ obeys the normalization condition $u^{\alpha}u_{\alpha}=-1$. For $\alpha>1$, the component of velocity becomes zero. Defining the radial component of the four-dimensional velocity as $\upsilon(r)=u^1=\frac{\mathrm{d}r}{\mathrm{d}s}$, we get	from the normalization condition
	\begin{eqnarray}
		\label{3}
		(u^0)^2=\frac{1-2M/r+\upsilon^2}{\left(1-2M/r\right)^2}.
	\end{eqnarray}
If neglecting the self-gravity of the flow, we can determine the baryon number density $n$ and the number flux density $J^{\alpha}=nu^{\alpha}$ in the flow's local inertial frame. Assuming no particle generation or disappearance, thus the number of particles conserves, which gives
		\begin{eqnarray}
			\label{4}
			\nabla_{\alpha}J^{\alpha}=\nabla_{\alpha}(nu^{\alpha})=0,
		\end{eqnarray}
were $\nabla_{\alpha}$ denotes the covariant derivative with respect to the spacetime coordinates. For the metric (\ref{1}), the equation (\ref{4}) can be rewritten as
		\begin{eqnarray}
			\label{6}
			\frac{1}{r(r-a)}\frac{\mathrm{d}r(r-a)n\upsilon}{\mathrm{d}r}=0,
		\end{eqnarray}
which for a perfect fluid gives the integration as
		\begin{eqnarray}
			\label{7}
			r(r-a)n\upsilon=C_1,
		\end{eqnarray}
where $C_1$ is an integration constant. Integration the equation (\ref{6}) over the spatial volume and multiplying with the mass of each particle, $m$, we can obtain
		\begin{eqnarray}
			\label{8}
			\dot{M}=4\pi r(r-a)mn\upsilon,
		\end{eqnarray}
where $\dot{M}$ is a constant of integration which has dimension of mass per unit time. Actually, this is the Bondi mass accretion rate. Comparing the equation \eqref{8} with the equation (40) in \cite{Yang:2021opo}, we find that these two equations have similar forms, but there are significant differences: at horizon the equation \eqref{8} reduces to $\dot{M}=8\pi M(2M-a)mn_{H}\upsilon_{H}$, obviously $n_{H}$ and $\upsilon_{H}$ are very different from $u^0_{\infty}$ and $n_\infty$ (the time component of velocity and the number density at infinity) in the equation (40) in \cite{Yang:2021opo}, respectively.
		
Considering the accretion of adiabatic fluids does not disturb the global spherical symmetry of the black hole, the $\beta=0$ component of the energy-momentum conservation, $\nabla_{\alpha}T^{\alpha}_{\beta}=0$, gives for steady-state flow
		\begin{eqnarray}
		\label{9}			
       \frac{1}{r(r-a)}\frac{\mathrm{d}r(r-a)\upsilon(\rho+p)\sqrt{1-\frac{2M}{r}+\upsilon^2}}{\mathrm{d}r}=0,
		\end{eqnarray}
which can be integrated as
		\begin{eqnarray}
			\label{10}
			r(r-a)\upsilon(\rho+p)\sqrt{1-\frac{2M}{r}+\upsilon^{2}}=C_{2},
		\end{eqnarray}
where $C_2$ is an integration constant. Dividing the equations (\ref{10}) and (\ref{7}) and the squaring, we acquire
\begin{eqnarray}
	\label{11} \left(\frac{\rho+p}{n}\right)^2\left(1-\frac{2M}{r}+\upsilon^2\right)=\left(\frac{\rho_{\infty}+p_{\infty}}{n_{\infty}}\right)^2.
\end{eqnarray}
The $\beta=1$ component of the energy-momentum conservation, $\nabla_{\alpha}T^{\alpha}_{\beta}$, gives
\begin{eqnarray}
	\label{12}	\upsilon\frac{\mathrm{d}\upsilon}{\mathrm{d}r}=\frac{1-\frac{2M}{r}+\upsilon^2}{\rho+p}\frac{\mathrm{d}p}{\mathrm{d}r}-\frac{1}{2}\frac{2M}{r^2}.
\end{eqnarray}
The equations (\ref{8}) and (\ref{11}) represent the fundamental conservation equations governing the material flow onto the charged dilaton black hole, neglecting the back-reaction of matter.

\section{CONDITIONS FOR CRITICAL ACCRETION}
In this section, we considered the conditions for critical accretion. In the local inertial rest frame of the fluid, the conservation of mass-energy for an adiabatic fluid without entropy generation is governed by
\begin{eqnarray}
	\label{13}
	0=T\mathrm{d}s=\mathrm{d}\frac{\rho}{n}+p\mathrm{d}\frac{1}{n},
\end{eqnarray}
from which we can obtain the following relationship
\begin{eqnarray}
	\label{14}
	\frac{\mathrm{d}\rho}{\mathrm{d}n}=\frac{\rho+p}{n}.
\end{eqnarray}
According to this equation, the adiabatic sound speed of the fluid is defined as
\begin{eqnarray}
	\label{15}
	c^2_{\rm s}\equiv\frac{\mathrm{d}p}{\mathrm{d}\rho}=\frac{n}{\rho+p}\frac{\mathrm{d}p}{\mathrm{d}n}.
\end{eqnarray}
Differentiating the equations (\ref{7}) and (\ref{12}) with respect to $r$, yields
\begin{eqnarray}
	\label{16}
	\frac{1}{\upsilon}\upsilon^{\prime}+\frac{1}{n}n^{\prime}=-\frac{2r-a}{r(r-a)},
\end{eqnarray}
\begin{eqnarray}
	\label{17}
	\upsilon\upsilon^{\prime}+\left(1-\frac{2M}{r}+\upsilon^2\right)\frac{c_{\rm s}^2}{n}n^{\prime}=-\frac{M}{r^2}.
\end{eqnarray}
We can express these equations as the following system
\begin{eqnarray}
	\label{18}
	\upsilon^{\prime}=\frac{N_1}{N},
\end{eqnarray}
\begin{eqnarray}
	\label{19}
	n^{\prime}=-\frac{N_2}{N},
\end{eqnarray}
where
\begin{eqnarray}
	\label{20}
	N_1=\frac{1}{n}\left[\frac{(2r-a)c_{\rm s}^2}{r(r-a)}\left(1-\frac{2M}{r}+\upsilon^2\right)-\frac{M}{r^2}\right],
\end{eqnarray}
\begin{eqnarray}
	\label{21}
	N_2=\frac{1}{\upsilon}\left[\frac{(2r-a)\upsilon^2}{r(r-a)}-\frac{M}{r^2}\right],
\end{eqnarray}
\begin{eqnarray}
	\label{22}
	N=\frac{\upsilon^2-(1-\frac{2M}{r}+\upsilon^2)c_{\rm s}^2}{u\upsilon}.
\end{eqnarray}
Considering large values of $r$, we can require the flow to be subsonic, which means that $\upsilon^2<c^2_{\rm s}$. Since the speed of sound is always less than the speed of light, $c_{\rm s}<1$, we have $\upsilon^2\ll1$. The equation (\ref{22}) can be approximated as
	\begin{eqnarray}
		\label{23}
		N\approx \frac{\upsilon^2-c_{\rm s}^2}{n\upsilon},
	\end{eqnarray}
Since $\upsilon<0$, we obtained $N>0$ as $r\rightarrow\infty$. At the event horizon, $r_H=2M$, we get
	\begin{eqnarray}
		\label{24}
		N=\frac{\upsilon^2[1-c_{\rm s}^2]}{n\upsilon},
	\end{eqnarray}	
In adherence to the causality constraint, $c_{\rm s}^2<1$ implies $N<0$. Hence, there must exist critical points $r_c$ where $r_H<r_c<\infty$, at which $N=0$. The flow must traverse these critical points outside the horizon to avoid discontinuity. To ensure a smooth transition, it is necessary to impose $N=N_1=N_2=0$ at $r=r_c$, namely
	\begin{eqnarray}
		\label{25}
		N_1=\frac{1}{n_{c}}\left[\frac{(2r_{c}-a)c_{\rm s}^{2}}{r_{c}(r_{c}-a)}\left(1-\frac{2M}{r_{c}}+c^{2}_{c}\right)-\frac{M}{r^{2}_{c}}\right]=0,
		\end{eqnarray}
		\begin{eqnarray}
			\label{26}
			N_2=\frac{1}{\upsilon_c}\left[\frac{(2r_c-a)\upsilon^{2}_{c}}{r_{c}(r_{c}-a)}-\frac{M}{r^{2}_{c}}\right]=0,
		\end{eqnarray}
		\begin{eqnarray}
			\label{27}
			N=\frac{\upsilon^{2}_{c}-(1-\frac{2M}{r_{c}}+\upsilon^{2}_{c})c_{c}^{2}}{\upsilon_{c}n_{c}}=0,
		\end{eqnarray}
where $\upsilon_{c}=\upsilon(r_{c})$ and $c^2_{c}=c^2(r_{c})$. From the equations (\ref{25}), (\ref{26}), and (\ref{27}), we can derive the radial velocity, the speed of sound at the critical points, and the critical points, respectively
		\begin{eqnarray}
			\label{28}
			\upsilon^{2}_{c}=\frac{M(r_{c}-a)}{r_{c}(2r_{c}-a)},
		\end{eqnarray}
		\begin{eqnarray}
			\label{29}
			c_{c}^{2}=\frac{\upsilon^{2}_{c}}{1-\frac{2M}{r_{c}}+\upsilon^{2}_{c}}.
		\end{eqnarray}
Only for $\upsilon^{2}_{c}\geq 0$ and $c_{c}^{2}\geq 0$, the equations (\ref{18}) and (\ref{19}) admit acceptable solutions. Therefore we have
		\begin{eqnarray}
			\label{32}
			r_{c}>\frac{3M+a+\sqrt{9M^{2}-2Ma+a^{2}}}{4}.
			\end{eqnarray}
This equation indicates the conditions that allow the equations (\ref{28}) and (\ref{29}) to have physically meaningful solutions. Taking into account the causality constraints $c_{c}^{2}\leq 1$, we infer $r_{c}\geq r_{H}=2M$, signifying that the critical points are situated outside the event horizon. Since $a\leq 2M$, so $(3M+a+\sqrt{9M^{2}-2Ma+a^{2}})/4\leq 2M$. Combining these two  inequalities, we determine the conditions satisfied by the critical points: $r_{c}\geq r_{H}=2M$.

			\section{THE POLYTROPIC SOLUTION}
In this section, we explored the accretion rate for a polytropic gas and calculated the gas compression as well as the adiabatic temperature distribution at the outer horizon.

			\subsection{Accretion for polytrope gas}
For explicit calculations, we adopted the polytrope gas introduced in \cite{hoyle_lyttleton_1939} and \cite{1940Obs6339L} with the equation of state
			\begin{eqnarray}
				\label{33}
				p=Kn^{\gamma},
			\end{eqnarray}
where $K$ is a constant and $\gamma$ is adiabatic index satisfying $1<\gamma<\frac{5}{3}$. By inserting this expression into the equation for the conservation of mass-energy (\ref{13}) and integrating, we can readily obtain		
			\begin{eqnarray}
				\label{34}
				\rho=\frac{K}{\gamma-1}n^{\gamma}+mn,
				\end{eqnarray}
where $mn$ is the rest energy density with $m$ an integration constant. With the definition of the speed of sound (\ref{15}), the Bernoulli equation (\ref{11}) can be rewritten as
			\begin{eqnarray}
				\label{35}
				\left(1+\frac{c_{\rm s}^{2}}{\gamma-1-c_{\rm s}^{2}}\right)^{2}\left(1-\frac{2M}{r}+\upsilon^{2}\right)=\left(1+\frac{c_{\infty}^{2}}
				{\gamma-1-c_{\infty}^{2}}\right)^{2}.
			\end{eqnarray}
Taking into account the critical radial velocity (\ref{28}) and the critical speed of sound (\ref{29}), the equation (\ref{35}) must satisfy the condition at the critical point $r_{c}$.
		 \begin{eqnarray}
				\label{36}
				\left(1-\frac{c_{c}^{2}}{\gamma-1}\right)^{2}\frac{8Mc_{c}^{2}+2M}{(4M-3a)c_{c}^{2}+2M}=\left(1-\frac{c_{\infty}^{2}}
				{\gamma-1}\right)^{2}.
			\end{eqnarray}
As $r_{c}<r<\infty$, the baryons are still non-relativistic, and we can anticipate that $c_{\rm s}^{2}<c_{c}^{2}\ll 1$. Expanding the equation (\ref{36}) to the first order in $c_{c}^{2}$ and $c_{\infty}^{2}$, results
			\begin{eqnarray}
				\label{37}
				c_{c}^{4}\approx \frac{4M c_{\infty}^{2}}{8M-3a-4M\gamma-3a\gamma}.
				\end{eqnarray}
According to this equation, we can obtain the expression of the critical points $r_{c}$ in terms of the boundary condition $c_{\infty}$ and the black-hole mass $M$ as
					\begin{eqnarray}
						\label{38}
						r_{c}\approx\frac{a+4M}{4}+\frac{8M+3a-4M\gamma-3a\gamma}{8c_{\infty}^{2}},
                    \end{eqnarray}						
From the equations (\ref{15}), (\ref{33}) and (\ref{34}), we get
						\begin{eqnarray}
							\label{39}
							\gamma Kn^{\gamma-1}=\frac{m c_{\rm s}^{2}}{1-\frac{c_{\rm s}^{2}}{\gamma-1}}.
						\end{eqnarray}
Since $c_{\rm s}^{2}\ll 1$, we have $n\sim c_{\rm s}^{\frac{2}{\gamma-1}}$
						\begin{eqnarray}
							\label{40}
							\frac{n_{c}}{n_{\infty}}=\left(\frac{c_{c}}{c_{\infty}}\right)^{\frac{2}{\gamma-1}}.
						\end{eqnarray}
Since $\dot{M}$ is a constant of integration, as implied by the equation (\ref{8}), it must also hold at $r=r_{c}$. With this condition, we can determine the accretion rate.
						\begin{eqnarray}
							\label{41}
							\dot{M}&=&4\pi r(r-a)mn\upsilon=4\pi r_{c}(r_{c}-a)mn_{c}\upsilon_{c} \nonumber\\
                           &=&4\pi A m n_{\infty},
                           \end{eqnarray}
                           where
						\begin{eqnarray}
							\label{45}
							A&=&\left(\frac{a+4M}{4}+\frac{8M-3a-4M\gamma-3a\gamma}{8c_{\infty}^{2}}\right)\left(\frac{4M-3a}{4}+\frac{8M-3a-4M\gamma-3a\gamma}
							{8c_{\infty}^{2}}\right)\nonumber\\
                          &\;&\left(\frac{4M}{8M-3a-4M\gamma-3a\gamma}\right)^{\frac{1}{\gamma-1}}
                          \left[\frac{4Mc_{\infty}^{2}\left(2c_{\infty}^{2}(4M-3a)+8M-3a-4M\gamma-3a\gamma\right)}{(8M-3a-4M\gamma-3a\gamma)(16Mc_{\infty}^{2}+8M-3a-4M\gamma-3a\gamma)}
                            \right]^{\frac{1}{2}}. \nonumber\\
							\;&\;&\;					
					\end{eqnarray}
Obviously, the dilaton parameter and the adiabatic index of accreted matter play important roles in the accretion process. For $r_{c}\leq r<\infty$, since $c_{\infty}\ll 1$, the parameter $A$ in the accretion rate (\ref{41}) reduces to
							\begin{equation}
							\begin{aligned}
								\label{42}
								A =\left(\frac{a+4M}{4}+\frac{8M-3a-4M\gamma-3a\gamma}{8c_{\infty}^{2}}\right)\left(\frac{4M-3a}{4}+\frac{8M-3a-4M\gamma-
						       3a\gamma}{8c_{\infty}^{2}}\right)\\
								\left(\frac{4M}{8M-3a-4M\gamma-3a\gamma}\right)^{\frac{1}{\gamma-1}}\left(\frac{4Mc_{\infty}^{2}}{8M-3a-4M\gamma-3a\gamma}
								\right)^{\frac{1}{2}},							
						\end{aligned}
					\end{equation}
which gives a constraint: $a<4M(2-\gamma)/[3(1+\gamma)]<2M$. This is a somewhat interesting result: the parameter $a$ is limited not only by the mass of black hole, but also by the adiabatic index of accreted material.

\begin{figure}
\centering
\includegraphics[height=7cm,width=10cm]{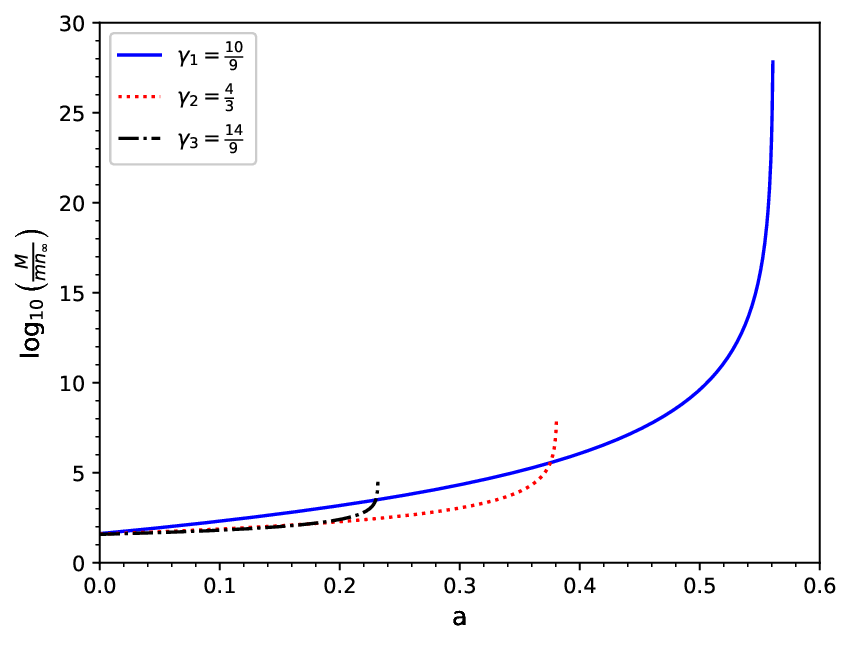}
\caption{Parametric $\frac{\dot{M}}{4\pi m n_{\infty}}-a$ diagrams with $M=1$, $c_{\infty}=0.01$, and $\gamma=10/9$, $\gamma=4/3$, $\gamma=14/9$, respectively.}
\label{m}
\end{figure}

In order to investigate the impact of the parameters on the accretion rate, we plot $\frac{\dot{M}}{4\pi m n_{\infty}}-a$ diagrams with some special values of the parameters in Fig. \ref{m}. We observe that the parameters $a$ and $\gamma$ have significant influences on the accretion rate, for example, it would approach the order $10^{30}$ for $\gamma=10/9$ and $a\simeq0.561$ and would become infinite if $a\rightarrow 4M(2-\gamma)/[3(1+\gamma)]$ for $M$ and $\gamma$ respectively taking a certain value within their allowed ranges.

\subsection{Asymptotic behavior at the event horizon}
Based on the findings from the preceding sections, the obtained results remain valid for large distances from the black hole, specifically in the vicinity of the critical radius $r_{c}\gg r_{H}$. Now we analysis the flow characteristics for $r_{H}<r\ll r_{c}$ and at the event horizon $r=r_{H}$.
						
For $r_{H}<r\ll r_{c}$, we obtained the fluid velocity from equation (\ref{35}), which can be approximated as
						\begin{eqnarray}
							\label{43}
							\upsilon^{2}\approx\frac{2M}{r}.
						\end{eqnarray}
At $r_{H}$, the flow speed approximately equals the speed of light, $\upsilon^{2}(r_{H})\approx 1$. Using the equations (\ref{8}), (\ref{41}), and (\ref{42}),we derive the gas compression at the horizon
						\begin{eqnarray}
							\label{44}
							\frac{n_{H}}{n_{\infty}}=\frac{A}{2M(2M-a)}.
						\end{eqnarray}
This equation also give a limit: $a<2M$, which is stronger than the limit coming from the equation (\ref{1}), but weaker than the limit coming from the inequality \eqref{42}.

\begin{figure}
\centering
\includegraphics[height=7cm,width=10cm]{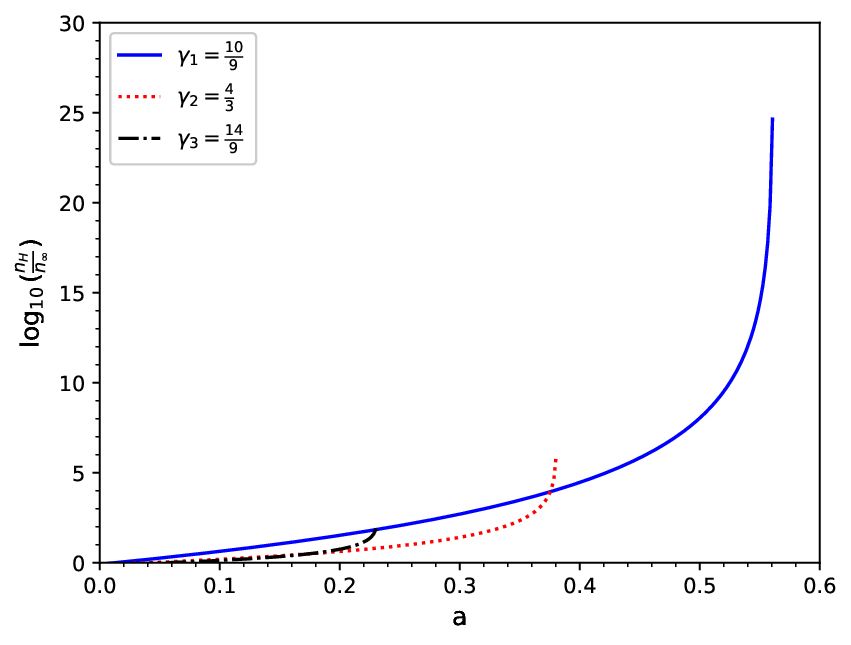}
\caption{Parametric $\frac{n_{H}}{n_{\infty}}-a$ diagrams with $M=1$, $c_{\infty}=0.01$, and $\gamma=10/9$, $\gamma=4/3$, $\gamma=14/9$, respectively.}
\label{n}
\end{figure}		
In Fig. \ref{n}, we plot $\frac{n_{H}}{n_{\infty}}-a$ diagrams with some special values of the parameters. We observe that the parameters $a$ and $\gamma$ also have significant impacts on the baryon number density which will be reduced by the mass of black hole.		
						
Employing the equations (\ref{33}) and (\ref{44}), we derive the adiabatic temperature profile at the event horizon for a Maxwell-Boltzmann gas with $p=n\kappa_{B}T$
						\begin{eqnarray}
							\label{46}
							\frac{T_{H}}{T_{\infty}}=\left(\frac{A}{4M^{2}}\right)^{\gamma-1}.
						\end{eqnarray}
Since $c_{\infty}\ll 1$, the parameter $A$ reduces to the equation (\ref{42}).						
\begin{figure}
\centering
\includegraphics[height=7cm,width=10cm]{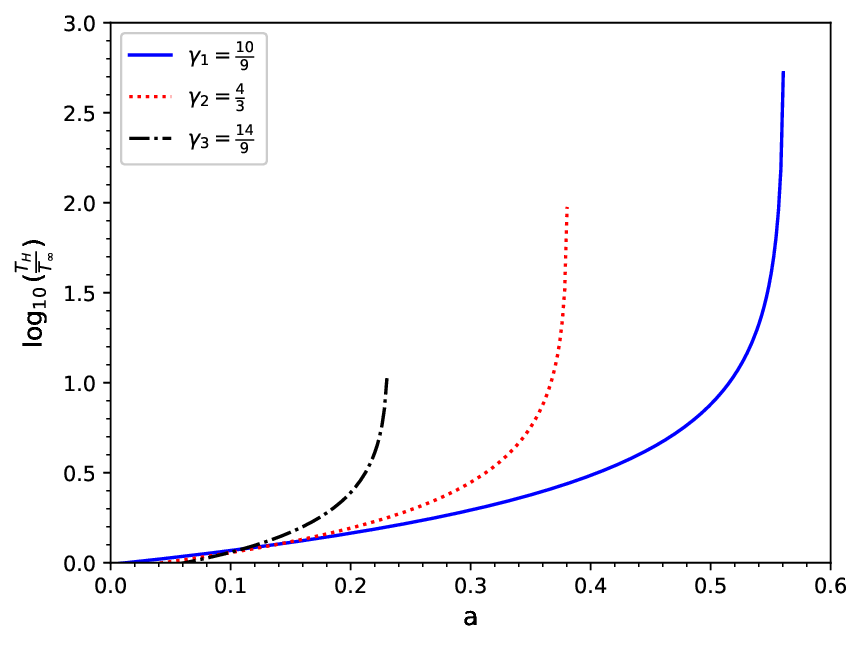}
\caption{Parametric $\frac{T_{H}}{T_{\infty}}-a$ diagrams with $M=1$, $c_{\infty}=0.01$, and $\gamma=10/9$, $\gamma=4/3$, $\gamma=14/9$, respectively.}
\label{T}
\end{figure}	

In Fig. \ref{T}, we plot $\frac{T_{H}}{T_{\infty}}-a$ diagrams with some special values of the parameters. We observe that the parameter $\gamma$ will reduce the magnitude of the temperature significantly, comparing with the mass accretion rate.					
						
						\section{CONCLUSIONS AND DISCUSSIONS}
In this study, we formulated and solved the problem of spherically symmetric, steady-state, adiabatic accretion onto a charged dilaton black hole within four-dimensional spacetime, employing general relativity. We derived fundamental equations for accretion and analytically determined the critical points, critical fluid velocity, critical sound speed, and subsequently the mass accretion rate. We identified the physical conditions that the critical points must satisfy. Explicit expressions for gas compression and temperature profiles were obtained below the critical radius and at the event horizon. We found that the dilaton parameter and the adiabatic index of accreted material have significant impacts on the accretion rate. We also discovered that the dilaton parameter is limited not only by the mass of black hole, but also by the adiabatic index of accreted material. Like the properties of thin accretion can give physical constraints on the parameter of Einstein-Aether-scalar theory \cite{He:2022lrc}, the process of accretion can provide new physical limits on the parameter of string theory; that is to say, in addition to theory and observation, physical processes can also give limits on the parameters of a model. We also compared the matter accretion rate with that obtained in \cite{Yang:2021opo} and pointed out their similarities and differences, which result is more reasonable, deserving further exploration. The outcomes here may offer valuable insights into comprehending the physical mechanisms governing accretion onto black holes.

						\begin{acknowledgments}
This study is supported in part by National Natural Science Foundation of China (Grant No. 12333008) and Hebei Provincial Natural Science Foundation of China (Grant No. A2021201034).
							
						\end{acknowledgments}

	\bibliographystyle{ieeetr}
\bibliography{acc}
					
				\end{document}